\documentclass[preprint,showpacs,preprintnumbers,amsmath,amssymb]{revtex4}

\usepackage{epsfig}

\begin{document}
\title{Innovation flow through social networks: Productivity distribution}

\author{T. Di Matteo} 
\email{tiziana.dimatteo@anu.edu.au; Tel ++ 61 (0)2 61250166; FAX ++ 61 (0)2 61250732.}
\affiliation{ Department of Applied Mathematics, Research School of Physical Sciences and Engineering, The Australian National University, Canberra ACT 0200, Australia.}
\author{T. Aste}
\affiliation{ Department of Applied Mathematics, Research School of Physical Sciences and Engineering, The Australian National University, Canberra ACT 0200, Australia.}

\author{M. Gallegati}
\affiliation{Department of Economics, Universit\`a Politecnica delle Marche, Piaz.le Martelli 8, I-60121 Ancona, Italy.}

\date{\today}

\begin{abstract}
A detailed empirical analysis of the productivity of non financial firms across several countries and years shows that productivity follows a non-Gaussian distribution with power law tails.
We demonstrate that these empirical findings can be interpreted as consequence of a mechanism of exchanges in a social network where firms improve their productivity by direct innovation or/and by imitation of other firm's technological and organizational solutions.
The type of network-connectivity determines how fast and how efficiently information can diffuse and how quickly innovation will permeate or behaviors will be imitated. 
From a model for innovation flow through a complex network we obtain that the expectation values of the productivity level are proportional to the connectivity of the network of links between firms.
The comparison with the empirical distributions reveals that such a network must be of a scale-free type with a power-law degree distribution in the large connectivity range.
\end{abstract}


\maketitle

\section{Introduction}

Recently, the availability of huge sets of longitudinal firm-level data has generated a soars of productivity studies in the economic literature \cite{Ijiri,Axtell,Gaffeo,Gibrat,Sutton,Barnes,Kruger}. There are several measures of productivity \cite{Hulten2000}, in this work we consider two basic measures: labour and capital productivity. The Labour productivity is defined as value added over the amount of employees (where value added, defined according to standard balance sheet reporting, is the difference between total revenue and cost of input excluding the cost of labour). Although elementary, this measure has the advantage of being accurately approximated given the available data. The other alternative measure is the capital productivity which is defined as the ratio between value added and fixed assets (i.e. capital). This second measure has some weakness since the firms' assets change continuously in time (consider for instance the value associated with the stock price). Usually the literature recognizes that the productivity distribution is not normally distributed \cite{Kruger}, and empirically `fat tails' with power law behaviors are observed.
But the mainstream proposed explanations cannot retrieve this power law tails yielding -at best- to log-normal distributions \cite{Hopenhayn,Ericson}.
According to the evolutionary perspective \cite{Nelson1982,Nelson1995}, firms improve their productivity implementing new technological and organizational solutions and, by this way, upgrading their routines. The search for more efficient technologies is carried out in two ways: (1) by {\it innovation} (direct search of more efficient routines); (2) by {\it imitation} of the most innovative firms \cite{Dosi,Mazzuccato}. 
In practice, one can figure out that once new ideas or innovative solutions are conceived by a given firm then they will percolate outside the firm that originally generated them by imitation from other firms. 
In this way the innovation flows through the firms.
Therefore, the network of contacts between firms which allows such a propagation must play a decisive role in the process.

In this paper we introduce a model for the production and flow of innovation in a complex network linking the firms.
We show that the resulting productivity distribution is shaped by the connectivity distribution of this network and in particular we demonstrate that power law tails emerge when the contact-network is of a scale-free type.
These theoretical finding are corroborated by a large empirical investigation based on the data set \emph{Amadeus}, which records data of over 6 million European firms from 1990 to 2002 \cite{newpaper}. 
A statistical analysis of such a data reveals that: (i) the productivity is power law distributed in the tail region; (ii) this result is robust to different measures of productivity (added value-capital and capital-labor ratios); and (iii) it is persistent over time and countries \cite{newpaper}. A comparison with the theoretical prediction reveals that the empirical data are well interpreted by assuming that the contact network is of scale-free type with power law tailed degree distributions.

The paper is organized as follows: Section~\ref{s.second} recalls the concept of social network; Section ~\ref{S.m} introduces the model supporting the technological distribution while Section~\ref{s.EMTH} describes the empirical findings. A conclusive section summarizes the main results.

\section{Contact networks in social systems}
\label{s.second}
Systems constituted of many elements can be naturally associated with networks linking interacting constituents. Examples in natural and artificial systems are: food webs, ecosystems, protein domains, Internet, power grids. In social systems, networks also emerge from the linkage of people or group of people with some pattern of contacts or interactions. Examples are: friendships between individuals, business relationships between companies, citations of scientific papers, intermarriages between families, sexual contacts. The relevance of the underlying connection-network arises when the collective dynamics of these systems is considered. Recently, the discovery that, above a certain degree of complexity, natural, artificial and social systems are typically characterized by networks with power-law distributions in the number of links per node (degree distribution), has attracted a great deal of scientific interest \cite{Barabasi,Newman,Amaral}. Such networks are commonly referred as scale-free networks and have degree distribution: $p_k \sim k^{-\alpha}$ (with $p_k$ the probability that a vertex in the network chosen uniformly at random has degree $k$). In scale-free networks most nodes have only a small number of links, but a significant number of nodes have a large number of links, and all frequencies of links in between these extremes are represented. The earliest published example of a scale-free network is probably the study of Price \cite{Price} for the network of citations between scientific papers. Price found that the exponent $\alpha$ has value $2.5$ (later he reported a more accurate figure of $\alpha=3.04$). More recently, power law degree distributions have been observed in several networks, including other citation networks, the World Wide Web, the Internet, metabolic networks, telephone calls and the networks of human sexual contacts \cite{Barabasi,Newman,Liljeros,Mossa,Gabor}. All theses systems have values of the exponents $\alpha$ in a range between 0.66 and 4, with most occurrences between $2$ and $3$ \cite{Bara04,Aleberich02,Bara4,Watts98}. 

When analyzing the industrial dynamics, it is quite natural to consider the firms as interacting within a network of contacts and communications. 
In particular, when the productivity is concerned, such a network is the structure through which firms can imitate each-other. Our approach mimics such a dynamics by considering simple type of interactions but assuming that they take place through a complex network of contacts.

\section{Innovation flow}
\label{S.m}

The innovation originally introduced in a given firm `$i$' at a certain time $t$ can spread by imitation across the network of contacts between firms. 
In this way, interactions force agents to progressively adapt to an ever changing environment. 

In this section we introduce a model for the flow of innovation through the system of firms.
 We start from the following equation describing the evolution in time of the productivity $x_l$ of a given firm `$l$':
\begin {eqnarray}
\label{W}
x_l(t+1)= x_l(t) + A_l(t)+ \sum_{j \in \mathcal{I}_l} Q_{j \to l}(t) [x_j(t) - x_j(t-1)] \\
-\sum_{\tau=l}^{t-1} q_{l}^{(\tau)}(t) [x_l(t-\tau) - x_l(t-\tau-1)] \nonumber.
\end {eqnarray}
The term $A_l(t)$ is a stochastic additive quantity which accounts the progresses in productivity due to innovation. The terms $Q_{j \to l}$ are instead exchange factors which model the imitation between firms. These terms take into account the improvement of the productivity of the firm '$l$' in consequence of the imitation of the processes and innovations that had improved the productivity of the firm '$j$' at a previous time. Such coefficients are in general smaller than one because the firms tend to protect their innovation content and therefore the imitation is -in general- incomplete. In the following we will consider only the static cases where these quantity are independent on $t$. 
The term $q_l^{(\tau)}$ is:
\begin{eqnarray}
\label{q}
q_l^{(1)} &=& \sum_{j \in \mathcal{I}_l} Q_{j \to l} Q_{l \to j} \;\; \mbox{for $\tau=1$} \\
q_l^{(\tau)} &=& \sum_{j \in \mathcal{I}_l} Q_{j \to l} \sum_{h_1 \ldots h_{\tau-1}} Q_{l \to h_1} Q_{h_1 \to h_2} \ldots Q_{h_{ {\tau-1} \to j}} \;\; \mbox{for $\tau \geq 2$}.
\end{eqnarray}
This term excludes back-propagation: firm `$l$' imitates only improvements of the productivity of firm `$j$' which have not been originated by imitation of improvements occurred at the firm `$l$' itself at some previous time. 
The system described by Equation~\ref{W} can be viewed as a system of self-avoiding random walkers with sources and traps.

The probability $P_{t+1}(y,l)dy$ that the firm $l$ at the time $t+1$ has a productivity between $y$ and $y+dy$ is related to the probabilities to have a set $\{Q_{j\to l} \}$ of interaction coefficients and a set of additive coefficients $\{A_l(t)\}$ such that a given distribution of productivity $\{x_j(t)\}$ at the time $t$ yields, through Equation~\ref{W}, to the quantity $y$ for the agent $l$ at time $t+1$. 
This is:
\begin{eqnarray}
\label{Pw-1}
P_{t+1}(y,l) &=&
\int_{-\infty}^\infty \ da \,
\Lambda_t(a,l) \prod_{\xi=0}^{t-1}
\int_{-\infty }^\infty  dx_1^{(\xi)} P_{t-\xi}(x_1^{(\xi)},1)
\cdots \\ \nonumber
&&
\int_{-\infty }^\infty  dx_N^{(\xi)} P_{t-\xi}(x_N^{(\xi)},N) \\ \nonumber
&&
\delta \big(y - a -  x_l^{(0)} - \sum_{j \in \mathcal{I}_l}  [x_{j}^{(0)}- x_{j}^{(1)}] Q_{j\to l} + \sum_{\tau=l}^{t-1} q_{l}^{(\tau)} [x_l^{(\tau)} - x_l^{(\tau+1)}]  \big) ,
\end{eqnarray}
where $\delta(y)$ is the Dirac delta function and $\Lambda_t(a,l)$ is the probability density to have at time $t$ on site $l$ an
additive coefficient $A_l(t)=a$. 
Let us introduce the Fourier transformation of $ P_t(y,l)$ and its inverse
\begin{eqnarray}
\label{FP}
\hat P_t(\varphi,l) &=&  \int_{-\infty}^\infty  dy
e^{+ i y \varphi} P_t(y,l)
\nonumber \\
     P_t(y,l)       &=&  \frac{1}{2\pi} \int_{-\infty}^\infty  d\varphi
e^{- i y \varphi}\hat P_t(\varphi,l) \;\;\;.
\end{eqnarray}
In appendix \ref{A}, we show that Equation~\ref{Pw-1} can be re-written in term of these transformations, resulting in: 
\begin{eqnarray}
\label{Pw5}
\hat P_{t+1}(\varphi,l)
&=&
\hat \Lambda_t(\varphi,l)
\hat P_t(\varphi,l)
\prod_{\xi=2}^{t-1}\hat P_{t-\xi}((-q_l^{(\xi)} +q_l^{(\xi-1)}) \varphi,l) \nonumber \\
&& \hat P_0(q_l^{(t-1)} \varphi,l) \hat P_{t-1}(-q_l^{(1)} \varphi,l)  \\
&& \prod_{j \in \mathcal{I}_l}
\hat P_t(Q_{j\to l} \varphi,j) \hat P_{t-1}(-Q_{j\to l} \varphi,j)\;\;, \nonumber
\end{eqnarray}
with $\hat \Lambda_t(\varphi,l)$ being the Fourier transform of $\Lambda_t(a,l)$.
From this equation we can construct a relation for the propagation of the cumulants of the productivity distribution. 
Indeed, by definition the cumulants of a probability distribution are given by the expression:
\begin{equation}
\label{C-1}
k^{(\nu)}_l(t) = (-i)^{\nu}\frac{d^{\nu}}{d\varphi^{\nu}} \ln \hat P_t(\varphi,l) \Big|_{\varphi=0} \;\;,
\end{equation}
where the first cumulant $k^{(1)}_l(t)$ is the expectation value of the stochastic variable $x_l$ at the time $t$ ($\left< x_l(t) \right> $) and the second cumulant $k^{(2)}_l(t)$ is its variance ($\sigma_l^2(t)$).
By taking the logarithm of Equation~\ref{Pw5} and applying Equation~\ref{C-1} we get:
\begin{eqnarray}
\label{C-2}
k^{(\nu)}_l(t+1) &=& c^{(\nu)} (t)
+ k^{(\nu)}_l(t) +\sum_{\xi=2}^{t-1} (q_l^{(\xi-1)} -q_l^{(\xi)})^{\nu} k^{(\nu)}_l(t-\xi) \nonumber \\
&& + (q_l^{(t-1)})^{\nu} k^{(\nu)}_l(0) +(-q_l^{(1)})^{\nu} k^{(\nu)}_l(t-1)
+ \\
&& \sum_{j \in \mathcal{I}_l} [\left(Q_{j\to l} \right)^\nu k^{(\nu)}_j(t) +\left(-Q_{j\to l} \right)^\nu k^{(\nu)}_j(t-1)] \;\;\;.\nonumber
\end{eqnarray}

It has been established by Maddison that the average innovation rate of change in the OECD countries since $1870$ has been roughly constant \cite{Maddison}. 
In our formalism this implies 
\begin{equation}
\frac{\left< A_l(t+1) \right> - \left< A_l(t) \right>  }{\left< A_l(t) \right>} \sim const.
\end{equation}
Therefore, the mean of the additive term in Equation~\ref{W} ($\left< A_l(t) \right>$) must grow exponentially with time and consequently the first cumulant (the average indeed) reads:  $c^{(1)}=c_0^{(1)} ( c_1^{(1)} )^t$. 
Equivalently we assume an exponential growth also for the other moments ($c^{(\nu)}=c_0^{(1)} ( c_1^{(\nu)} )^t$). 

Equation~\ref{C-2} can now be solved by using a mean-field, self-consistent solution (neglecting correlations and fluctuations in the interacting firms) obtaining:
\begin{eqnarray}
\label{S1}
k^{(1)}_l(t) &=& \frac{1}{A}\frac{c_0^{(1)}  c_1^{(1)} }{(c_1^{(1)} -1)} \Big [1 + {\bar a }  Q z_l \Big ] (c_1^{(1)} )^{t} \nonumber \;\; \;\; \;\; \;\; \mbox{for $\nu=1$}
\\
k^{(\nu)}_l(t) &=&
\frac{c_0^{(\nu)} }{B_{\nu}} \Big [1 + (1 + \frac{(-1)^{\nu}}{c_1^{(\nu)}}) {\bar b^{(\nu)} }  Q^{\nu} z_l \Big ] (c_1^{(\nu)} )^t  \;\; \;\; \;\;\;\;\mbox{for $\nu>1$}
\end{eqnarray}
where
\begin{eqnarray}
\label{abdf}
&& {\bar a} =  \frac{1}{1 - \left<\frac{Q z_l}{A}\right>} \frac{1}{\left< A\right >} \\
&& {\bar b^{(\nu)} } =  \frac{1}{1 + \left<\frac{(1 + (-1)^{\nu}/c_1^{(\nu)}) Q^{\nu} z_l}{B_{\nu}}\right>} \frac{1}{\left< B_{\nu}\right>}
\end{eqnarray}
and
\begin{eqnarray}
\label{A1B1AvBv}
A &=& c_1^{(1)}  + z_l \sum_{\xi=1}^{t-1} \frac{Q^{\xi+1}} {(c_1^{(1)} )^{\xi}} \\
B_{\nu} &=&  -1 + c_1 ^{(\nu)} - z_l^{\nu} \big[\frac{(-Q^2)^{\nu}}{c_1^{(\nu)}} \nonumber \\
&+& \sum_{\xi=2}^{t-1} \frac{(Q^{\xi} - Q^{\xi+1})^{\nu}}{(c_1^{(\nu)} )^{\xi}} + \frac{(Q^t)^{\nu}}{(c_1^{(\nu)} )^t} \big]
\end{eqnarray}
with $Q$ being the average exchange factor.
When this exchange term is small, Equation~\ref{S1} can be highly simplified by taking the first order in $Q$ only, leading to:
\begin{eqnarray}
\label{TT1}
k^{(1)}_l(t) &\sim& \frac{c_0^{(1)}}{c_1^{(1)}-1}
\Big[1+ z_l \frac{Q}{c_1^{(1)}}\Big](c_1^{(1)})^{t}  \nonumber  \\
k^{(\nu)}_l(t) &\sim& \frac{c_0^{(\nu)}}{c_1^{(\nu)}-1} (c_1^{(\nu)})^{t}
\end{eqnarray}
Equation~\ref{S1} (and its simplified form (Equation~\ref{TT1})) describes a mean productivity which grows at the same rate of the mean innovation growth (as a power of $c_1^{(1)}$) and is directly proportional to the number of connections that the firm has in the exchange network. From Equation~\ref{S1} we also have that all the cumulants increase with a corresponding power rate ($(c_1^{(\nu)})^t$). 
But, if we analyze the \emph{normalized cumulants}: $\lambda^{(\nu)}(t) = k^{(\nu)}_l(t)/[k^{(2)}_l(t)]^{\nu/2}$ we immediately see that at large $t$ they all tend to zero excepted for the mean and the variance. 
Therefore the probability distributions tend to Gaussians at large times.

Summarizing, in this section we have shown that, at large $t$, the expectation value of the productivity level of a given firm is proportional to its connectivity in the network of interaction and the fluctuations around this expectation-value are normally distributed. 
Each firm has a different connectivity and therefore the probability distribution for the productivity of the ensemble of firms is given by a normalized sum of Gaussians with averages distributed according with the network connectivity. 
As discussed in the previous section, power-law-tailed degree distributions are very common in many social and artificial networks. 
It is therefore natural to hypotheses that also the social/information network through which firms can exchange and imitate productivity has a degree distribution characterized by a power law in the large connection-numbers region. 
If this is the case, then the whole productivity distribution will show a power-law tail characterized by the same exponent of the degree distribution \cite{footnote1}. 

\section{Empirical analysis and comparison with theory}
\label{s.EMTH}
\begin{figure}[]
\begin{center}
\begin{tabular}{cc}
\mbox{\epsfig{file=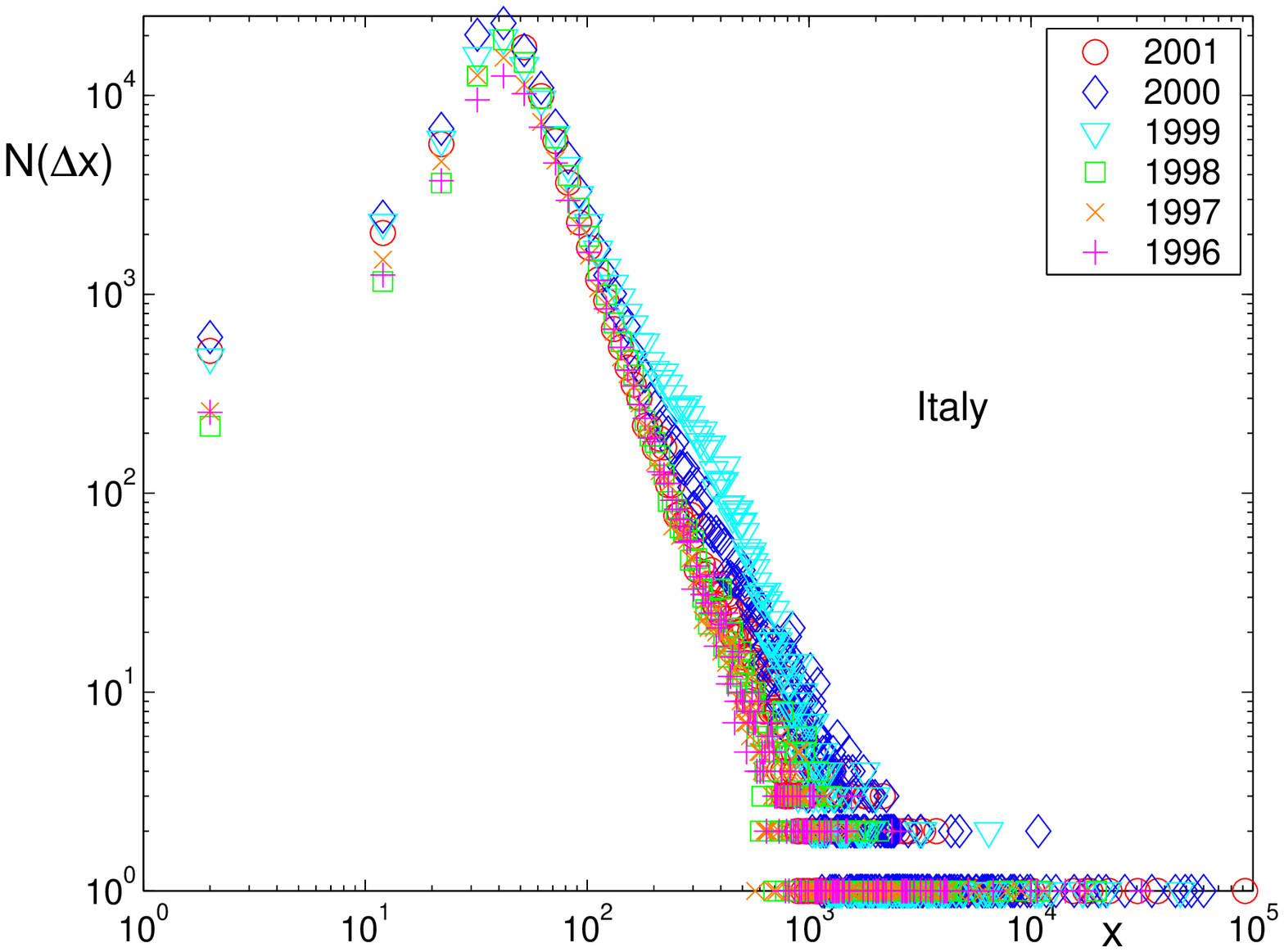,width=7.cm,height=5.cm,angle=0}}
&\mbox{\epsfig{file=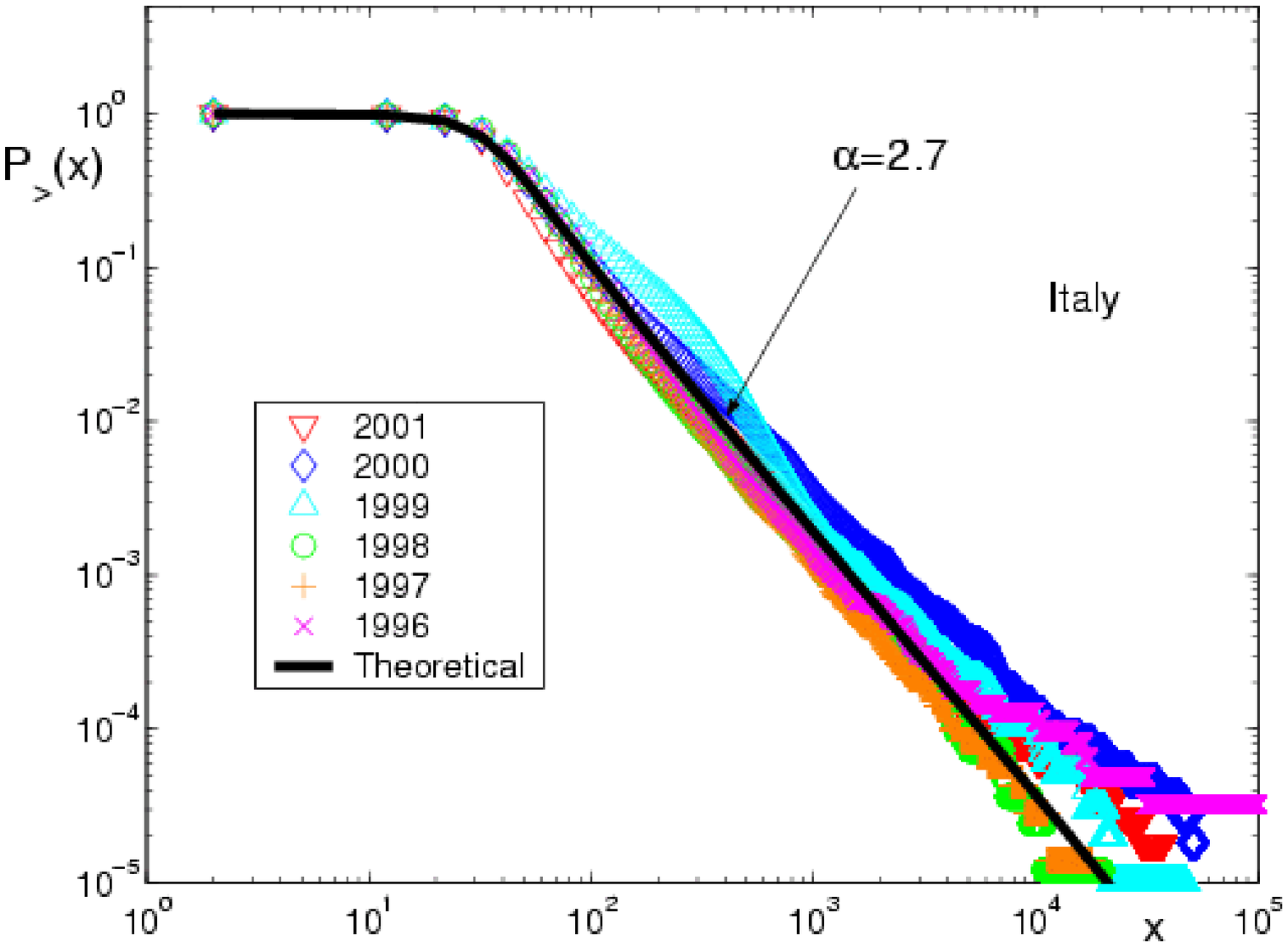,width=7.cm,height=5.cm,angle=0}}
\end{tabular}
\caption{Frequency distributions (Left) and complementary cumulative distributions (Right) for the labour productivity in Italy in the years $1996$-$2001$. The theoretical behavior is for $\alpha =2.7$, $m =22$, $n= 11$, $\sigma=10$ and $\beta =3$.}
\label{f.P2}
\end{center}
\end{figure}

\begin{figure}[]
\begin{center}
\begin{tabular}{cc}
\mbox{\epsfig{file=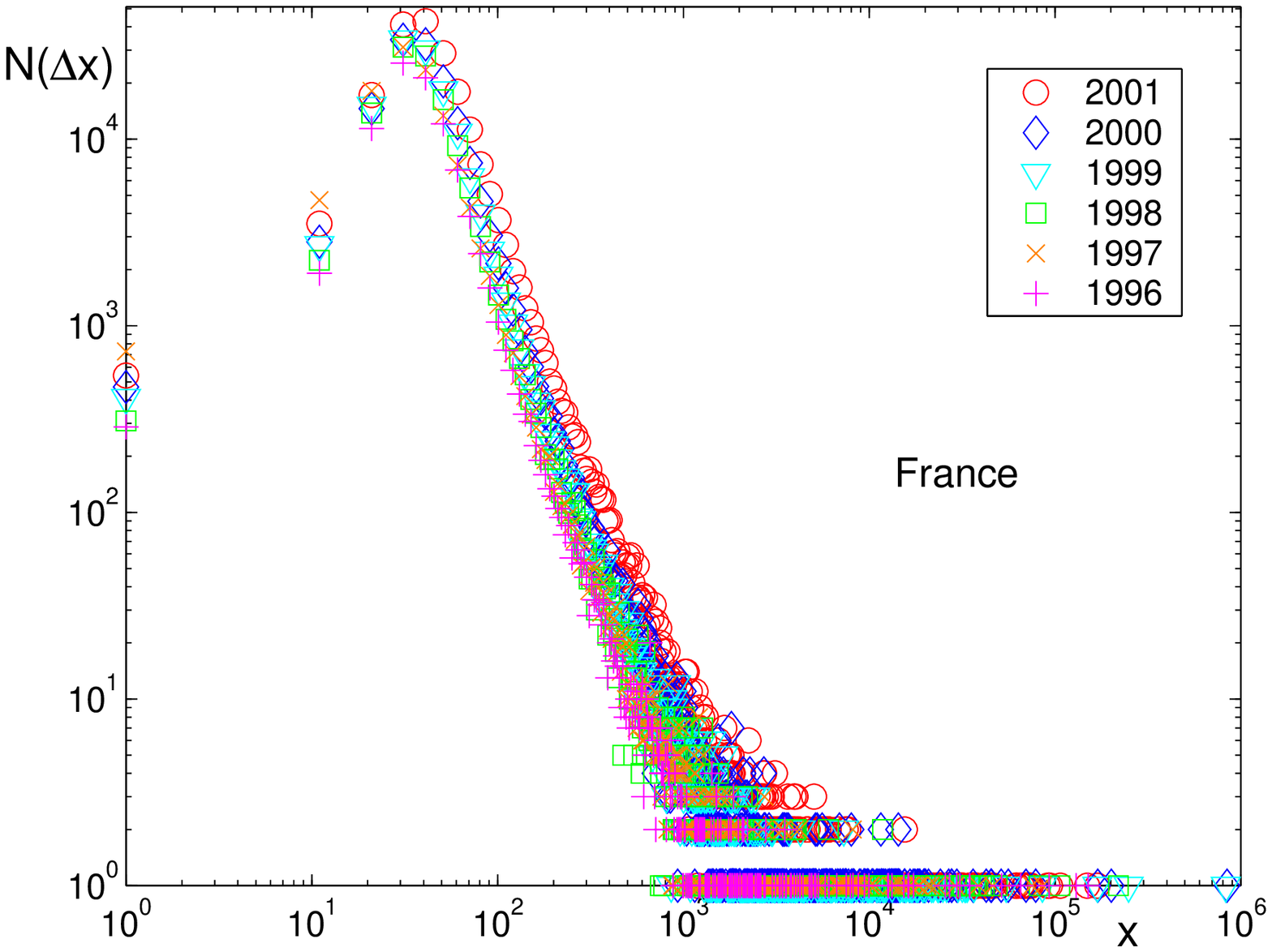,width=7.cm,height=5.cm,angle=0}}
&\mbox{\epsfig{file=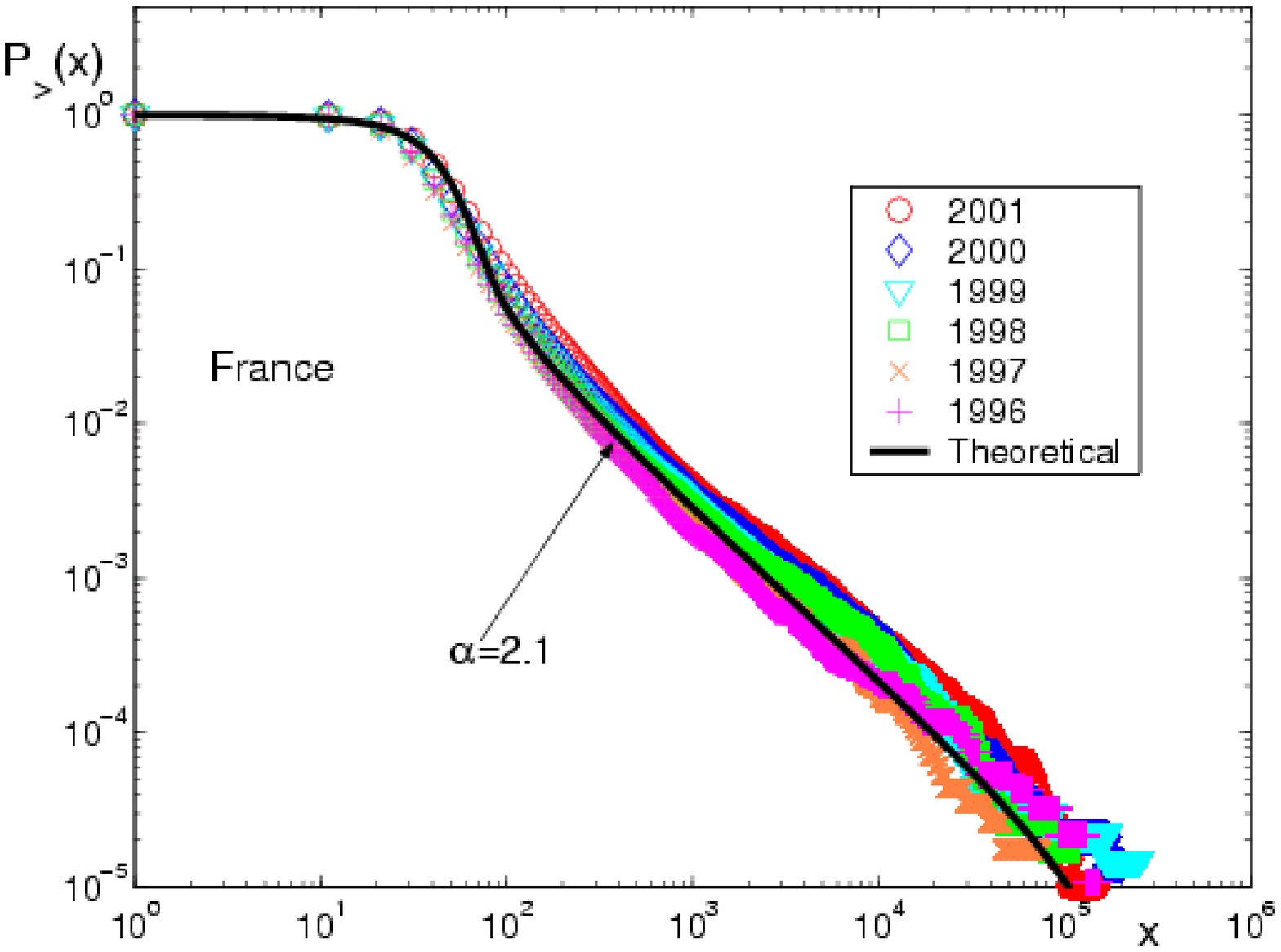,width=7.cm,height=5.cm,angle=0}}
\end{tabular}
\caption{Frequency distributions (Left) and complementary cumulative distributions (Right) for the labour productivity in France in the years $1996$-$2001$. The theoretical behavior is for $\alpha =2.1$, $m =30$, $n=4$, $\sigma=20$ and $\beta =1$.}
\label{f.P4}
\end{center}
\end{figure}

\begin{figure}[]
\begin{center}
\begin{tabular}{cc}
\mbox{\epsfig{file=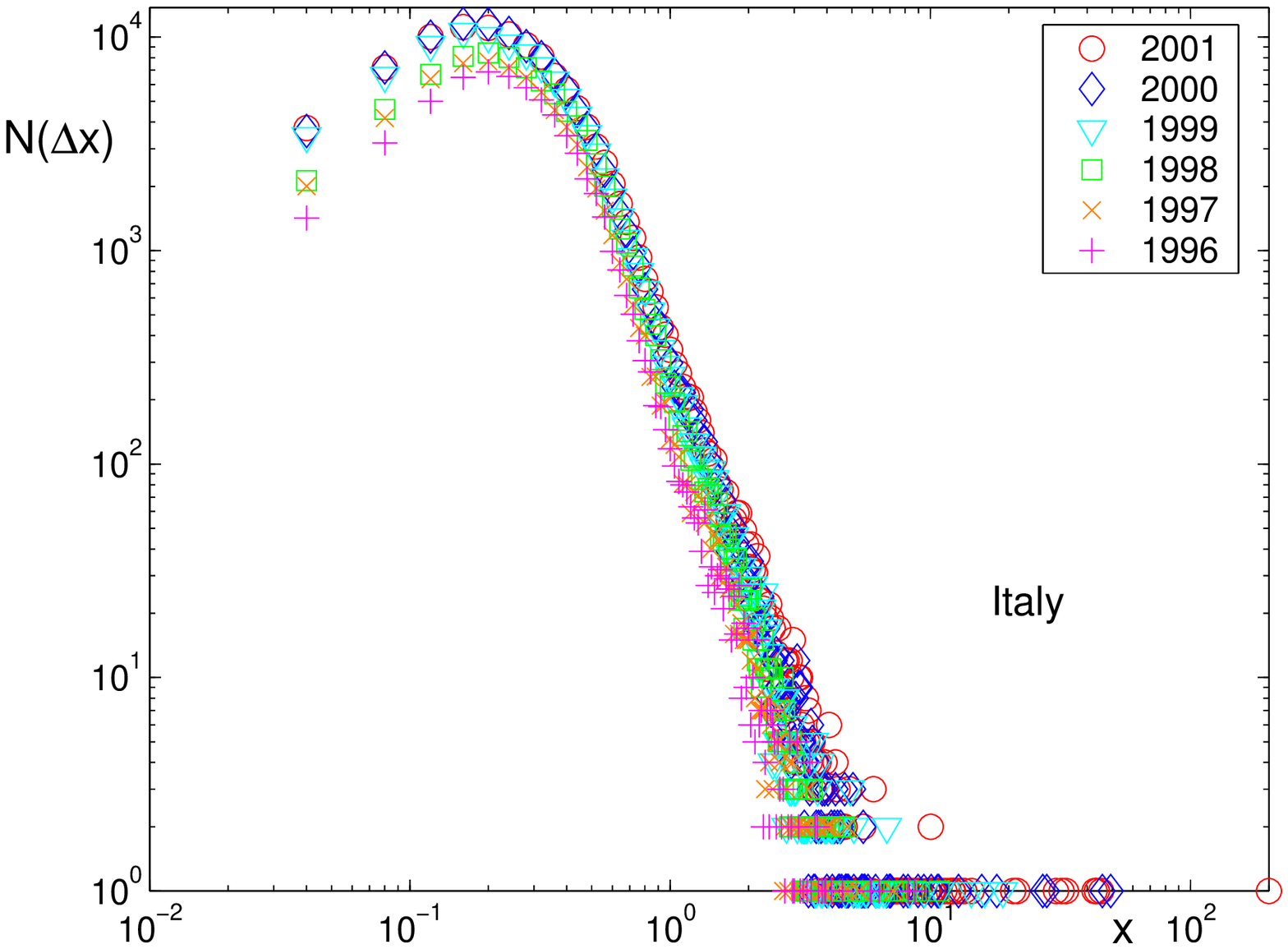,width=7.cm,height=5.cm,angle=0}}
&\mbox{\epsfig{file=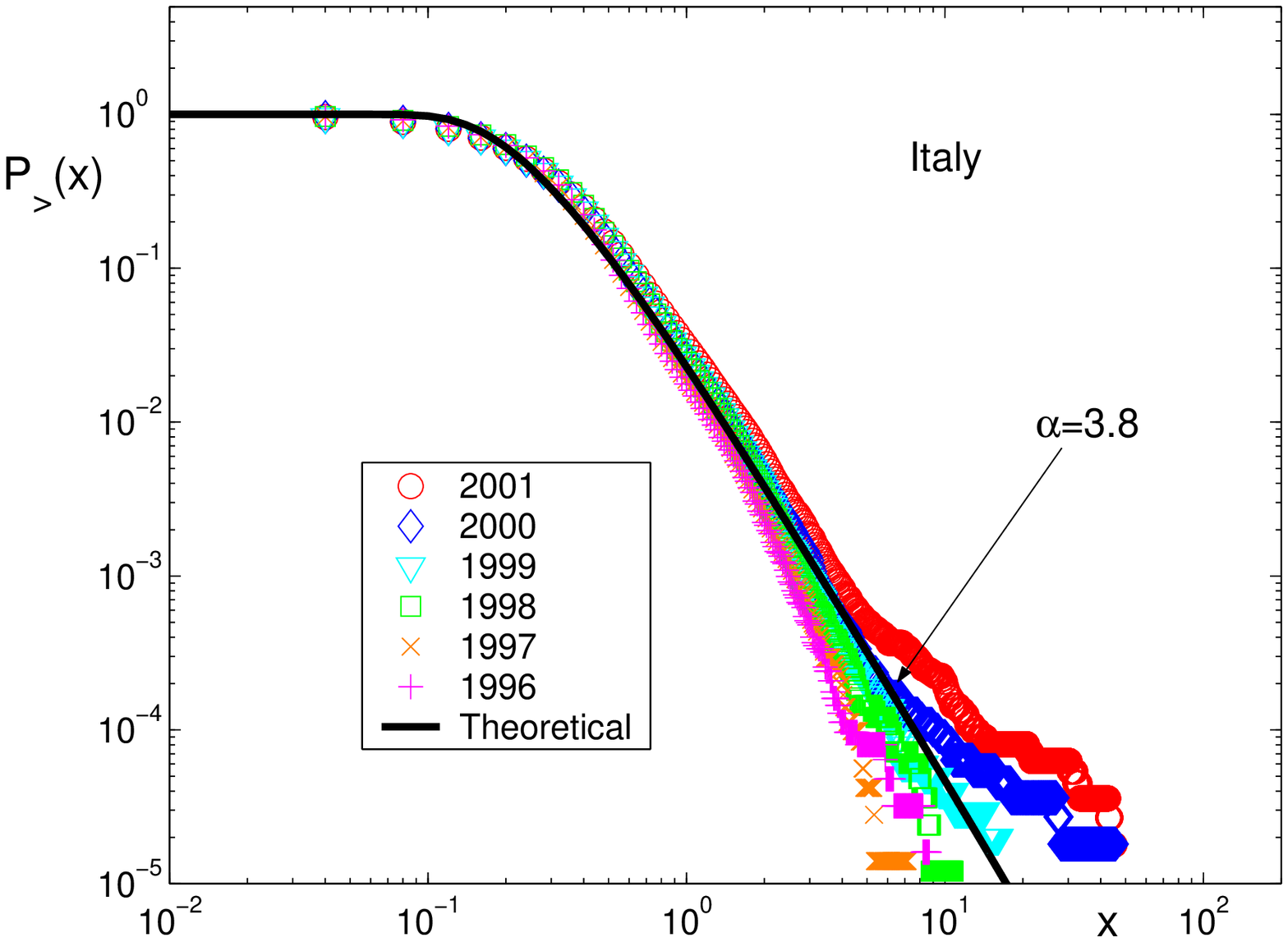,width=7.cm,height=5.cm,angle=0}}
\end{tabular}
\end{center}
\caption{Frequency distributions (Left) and complementary cumulative distributions (Right) for the capital productivity in Italy in the years $1996$-$2001$. The theoretical behavior is for $\alpha =3.8$, $m =0.04$, $n=0.02$, $\sigma=0.01$ and $\beta=25$.}
\label{f.P1}
\end{figure}

\begin{figure}[]
\begin{center}
\begin{tabular}{cc}
\mbox{\epsfig{file=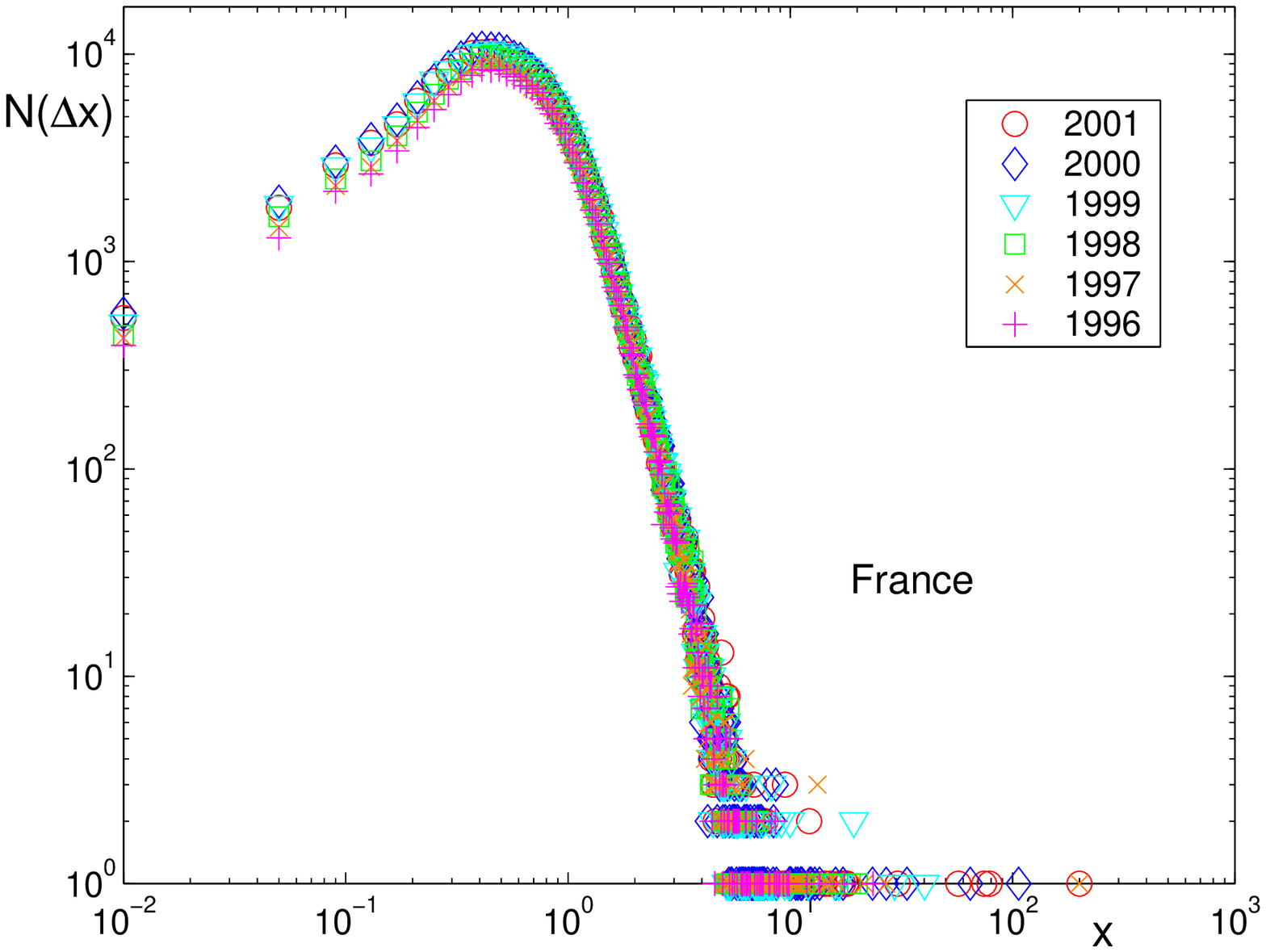,width=7.cm,height=5.cm,angle=0}}
&\mbox{\epsfig{file=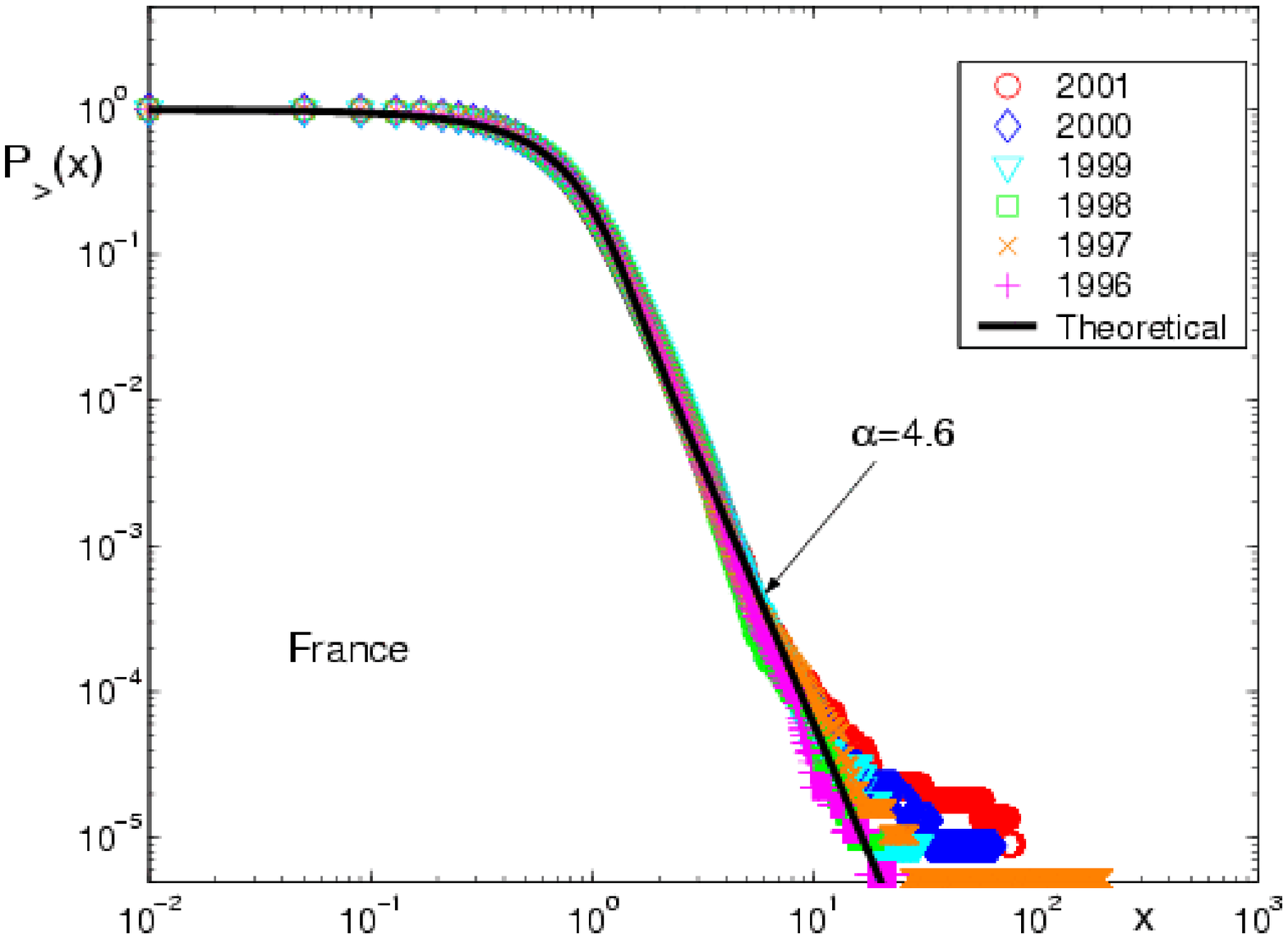,width=7.cm,height=5.cm,angle=0}}
\end{tabular}
\caption{Frequency distributions (Left) and complementary cumulative distributions (Right) for the capital productivity in France in the years $1996$-$2001$. The theoretical behavior is for $\alpha =4.6$, $m =0.06$, $n=0.02$, $\sigma=0.4$ and $\beta=68$.}
\label{f.P3}
\end{center}
\end{figure}

Figures~\ref{f.P2}, \ref{f.P4}, \ref{f.P1} and \ref{f.P3} show the log-log plot of the frequency distributions (Left) and the complementary cumulative distributions (Right) of labour productivity  and for capital productivity measured as quotas of total added value of the firms. In these figures the different data sets correspond to different years: $1996-2001$. For the sake of exposition, we illustrate the productivity distribution for France and Italy only, but similar results have been obtained for other Euroland countries of the AMADEUS dataset. The frequency distributions show a very clear non-Gaussian character: they are skewed with asymmetric tails and the labour productivity (Figures~\ref{f.P2} and ~\ref{f.P4} (Left)) present a clear leptokurtic pick around the mode. 
The complementary cumulative distributions ($P_>(x)$, being the probability to find a firm with productivity larger than $x$) show a linear trend at large $x$ implying a non-Gaussian character with the probability for large productivities well mimicked by a power-law behavior.

The model presented in this paper gives a simple explanation for the occurrence of such power law tails in the productivity distribution: they are a consequence of the social/information network which is of ``scale-free'' type (analogously with several other complex systems where such a connectivity-distribution can be measured \cite{DiMatteo,Richmond2001,Stanley1998,Biham1998,Bouchaud2000,Solomon2002}). Indeed, we have shown that distribution for the productivity of the ensemble of firms is given by a normalized sum of Gaussians with averages distributed according with the network connectivity. 
As consequence, when the connection network is of scale-free type the productivity distribution must share with it the same exponent in the power-law-tail. 

Comparisons between the theoretical predictions from Equation~\ref{TT1} associated with a scale-free network and the empirical findings are shown in the Figures~\ref{f.P2}, \ref{f.P4}, \ref{f.P1} and \ref{f.P3} (Right). 
In particular, accordingly with Equation~\ref{TT1}, we assume an average productivity given by $k^{(1)}_l = m +  z_l n$, a variance equal to $\sigma$ and the degree distribution of the network given by  $ p_k \propto k^{-\alpha} \exp(-\beta/k) $. The agreement with the empirical findings is quantitatively rather good.
We note that, although there are several parameters, the behavior for large productivity is controlled only by the power-law exponent $-\alpha$. 
On the other hand, in the small and the middle range of the distribution the other parameters have a larger influence.

From our analysis we observe that the theoretical curves fit well the empirical findings by assuming the power law exponent equal to $\alpha=2.7$ and $2.1$ for the labour productivity in Italy and France respectively. These exponents are in good agreement with the ones typical of the degree distribution in social networks. On the other hand the capital productivity presents much steeper decays which can be fitted with exponents $3.8$ and $4.6$ respectively. These very high values of the exponents might be consequence of the \textit{irrational euphoria} of the late $90$es when the stock markets were hit by a speculative bubble ($1997$) and its subsequent crash ($2000$). The bubble increased the value of the firms' asset thus lowering the value added-capital (i.e. capital productivity) ratio and soaring the power law coefficient of the power law distribution of the capital productivity distribution. However the very high capital productivity regions show a slowing down which could be fitted with lower exponents.

\section{Conclusions}

In this paper we have shown that the productivity of non-financial firms is power law distributed. This result is robust to different measures of productivity, different industrial sectors, years and countries. 
We have also argued that the empirical evidence corroborates the prescription of the evolutionary approach to technical change and demonstrated that power law distributions in productivity can be interpreted as consequence of a simple mechanism of exchanges within a social network. 
In particular, we have shown that the expectation values of the productivity level are proportional to the connectivity of the network of links between firms. 
The comparison with the empirical data indicates that such a network is of a scale-free type with a power-law degree distribution. 
In the present formulation we have assumed an underlying network which is fixed in time. 
This allows obtaining equilibrium solutions. 
On the other hand, a more realistic analysis should consider a non-static underlying network and therefore non-equilibrium trajectories modulated by the fluctuation in the underlying network. 
This non-equilibrium dynamics can be studied numerically from Equation~\ref{W} by using fluctuating exchange coefficients $Q_{j \to l}(t) $ . 
This is left to future research. In this paper we had a narrower goal: to show that empirical evidence is very well fitted by the evolutionary view of technical change.

\begin{acknowledgments}
We thank Corrado Di Guilmi for excellent research assistance. T. Di Matteo benefited from discussions with the participants to the COST P10 `Physics of Risk' meeting in Nyborg (DK), April 2004. TDM and TA acknowledge partially financial support from ARC Discovery project DP0344004 (2003).

\end{acknowledgments}

\appendix
\section{Cumulant propagation}
\label{A}

By using the Fourier transformation (Equation~\ref{FP}), Equation~\ref{Pw-1} becomes:
\begin{eqnarray}
\label{Pw1}
P_{t+1}(y,l)
&=&
\int_{-\infty}^\infty   da \Big\{\Lambda_t(a,l)
\prod_{\xi=0}^{t-1} \Big[
\frac{1}{(2\pi)^N}
\int_{-\infty}^\infty dx_1^{(\xi)} \cdots \int_{-\infty}^\infty dx_N^{(\xi)}
\\
&&
\int_{-\infty}^\infty   d\varphi_1^{(\xi)} e^{- i x_1^{(\xi)} \varphi_1^{(\xi)}}
\hat P_{t-\xi}(\varphi_1^{(\xi)},1)
\cdots \int_{-\infty}^\infty   d\varphi_N^{(\xi)} e^{- i x_N^{(\xi)} \varphi_N^{(\xi)}}
\hat P_{t-\xi}(\varphi_N^{(\xi)},N) \Big]
\nonumber \\
&&
\frac{1}{2\pi}
\int_{-\infty}^\infty d\phi
e^{- i ( y - a - x_l^{(0)} - \sum_{j \in \mathcal{I}_l}  [x_{j}^{(0)}- x_{j}^{(1)}] Q_{j\to l} + \sum_{\tau=l}^{t-1} q_{l}^{(\tau)} [x_l^{(\tau)} - x_l^{(\tau+1)}] )\phi} \Big\},\nonumber
\end{eqnarray}
where the Dirac delta function has been written as
\begin{equation}
\label{dir}
\delta(y-y_0) = \frac{1}{2\pi}
\int_{-\infty}^\infty    d\phi e^{- i (y -y_0) \phi} \;\;\;.
\end{equation}
Equation~\ref{Pw1} can be re-written as:
\begin{eqnarray}
\label{Pw2}
&& P_{t+1}(y,l) =
\frac{1}{(2\pi)}
\int_{-\infty}^\infty    da \Big\{ \Lambda_t(a,l)
\int_{-\infty}^\infty    d\phi    e^{- i (y-a) \phi} \\
&&
\prod_{\xi=0}^{t-1} \Big[
\frac{1}{(2\pi)^N}
\int_{-\infty}^\infty   d\varphi_l^{(\xi)}
\Big(\hat P_{t-\xi}(\varphi_l^{(\xi)},l)
\int_{-\infty}^\infty   dx_l^{(\xi)}
e^{- i ( \varphi_l^{(0)} - \phi) x_l^{(0)}} e^{- i \sum_{\tau=2}^{t-1} ( \varphi_l^{(\tau)} + q_l^{(\tau)} \phi - q_l^{(\tau-1)} \phi) x_l^{(\tau)}} \nonumber \\
&& e^{- i ( \varphi_l^{(t)} - q_l^{(t-1)} \phi) x_l^{(t)}} e^{- i ( \varphi_l^{(1)} - q_l^{(1)} \phi) x_l^{(1)}} \Big) \prod_{j \in \mathcal{I}_l}
\int_{-\infty}^\infty   d\varphi_j^{(\xi)}
\Big(\hat P_{t-\xi}(\varphi_j^{(\xi)},j) \nonumber \\
&& \int_{-\infty}^\infty    dx_j^{(\xi)}
e^{ - i [( \varphi_j^{(0)} - Q_{j\to l}  \phi) x_j^{(0)} +( \varphi_j^{(1)} + Q_{j\to l}  \phi) x_j^{(1)}]}
\Big) \Big]\Big\}.
\nonumber
\end{eqnarray}
The integration over the $x$'s yields
\begin{eqnarray}
\label{Pw3}
&& P_{t+1}(y,l) =
\frac{1}{2\pi}
\int_{-\infty}^\infty   da \Big\{\Lambda_t(a,l)
\int_{-\infty}^\infty   d\phi \Big[
e^{- i (y-a) \phi }
\hat P_t(\phi,l) \\
&&\prod_{\xi=2}^{t-1}\hat P_{t-\xi}((-q_l^{(\xi)} +q_l^{(\xi-1)}) \phi,l) \hat P_0(q_q^{(t-1)} \phi,l) \hat P_{t-1}(-q_l^{(1)}\phi,l) \nonumber \\
&& \prod_{j \in \mathcal{I}_l}
\hat P_t(Q_{j\to l} \phi,j) \hat P_{t-1}(-Q_{j\to l} \phi,j)
\Big]\Big\}.\nonumber
\end{eqnarray}
Its Fourier transform is:
\begin{eqnarray}
\label{Pw4}
\hat P_{t+1}(\varphi,l)
&=&
\frac{1}{2\pi}
\int_{-\infty}^\infty    da \Big\{\Lambda_t(a,l)
\int_{-\infty}^\infty    d\phi \Big[ e^{ i a \phi}
\int_{-\infty}^\infty    d y e^{- i y (\phi - \varphi)}\\
&&
\hat P_t(\phi,l) \prod_{\xi=2}^{t-1}\hat P_{t-\xi}((-q_l^{(\xi)} +q_l^{(\xi-1)}) \phi,l) \hat P_0(q_q^{(t-1)} \phi,l) \hat P_{t-1}(-q_l^{(1)}\phi,l)\big] \nonumber \\
&& \prod_{j \in \mathcal{I}_l}
\hat P_t(Q_{j\to l} \phi,j) \hat P_{t-1}(-Q_{j\to l} \phi,j)
\Big\}.
\nonumber
\end{eqnarray}
Equation~\ref{Pw4} can be integrated over $y$ giving the Fourier transform of Equation~\ref{Pw-1} which is Equation~\ref{Pw5} in Section~\ref{S.m}.

\end{document}